 \documentclass[final,1p,times]{elsarticle}

\usepackage{amssymb}
 \usepackage{amsthm}

\usepackage{color}

\usepackage{subcaption}
\usepackage{here}

\journal{Physics letter A}

\begin{document}
\begin{frontmatter}



\title{Reduction of quantum noise using the quantum locking with an optical spring for gravitational wave detectors}


\author[label1]{Rika Yamada}
\author[label2]{Yutaro Enomoto}
\author[label1]{ Izumi Watanabe}
\author[label3]{ Koji Nagano}
\author[label4]{Yuta Michimura}
\author[label5]{Atsushi Nishizawa}
\author[label3]{ Kentaro Komori}
\author[label1]{ Takeo Naito}
\author[label1]{ Taigen Morimoto}
\author[label1]{ Shoki Iwaguchi}
\author[label1]{ Tomohiro Ishikawa}
\author[label4]{ Masaki Ando}
\author[label2]{ Akira Furusawa }
\author[label1]{ Seiji Kawamura}

\address[label1]{Department of Physics, Nagoya University, Furo-cho, Chikusa-ku, Nagoya, Aichi 464-8602, Japan}
\address[label2]{Department of Applied Physics, School of Engineering, The University of Tokyo, 7-3-1 Hongo, Bunkyo-ku, Tokyo 113-8656, Japan}
\address[label3]{Institute of Space and Astronautical Science, Japan Aerospace Exploration Agency, Sagamihara,
Kanagawa 252-5210, Japan}
\address[label4]{Department of Physics, The University of Tokyo, Bunkyo, Tokyo 113-0033, Japan}
\address[label5]{Research Center for the Early Universe (RESCEU), School of Science, The University of Tokyo, Tokyo 113-0033, Japan}

\begin{abstract}
In our previous research, simulation showed that a quantum locking scheme with homodyne detection in sub-cavities is effective in surpassing the quantum noise limit for Deci-hertz Interferometer Gravitational Wave Observatory (DECIGO) in a limited frequency range. This time we have simulated an optical spring effect in the sub-cavities of the quantum locking scheme. We found that the optimized total quantum noise is reduced in a broader frequency band, compared to the case without the optical spring effect significantly improving the sensitivity of DECIGO to the primordial gravitational waves.

\end{abstract}










\end{frontmatter}



\section{Introduction}
\label{intro}
In the latest observing run, Advanced LIGO \cite{aLIGO} and Advanced Virgo \cite{Virgo} had been detecting gravitational-wave signals from black hole/neutron star binaries at an average frequency of once or twice a week\cite{GWTC-2}. Recently, KAGRA \cite{KAGRA} also began observation and will join the LIGO and Virgo network shortly. However, gravitational-wave signals at low frequencies, especially bellow 10 Hz, are difficult to detect by the ground-based detectors because of ground vibration and thermal noise in the mirror suspensions. Thus it is expected that space-borne detectors are superior at low frequencies, as they are free from ground vibration and pendulum-like suspension. 

Primordial gravitational waves, which are expected to be produced during the inflation period, are among  the most important targets of low-frequency gravitational wave observation \cite{primordialGW}. Unfortunately, they have never been detected. To detect the primordial gravitational waves in addition to other important science goals, a Japanese space mission, Deci-hertz Interferometer Gravitational Wave Observatory (DECIGO), has been planned \cite{decigo, decigo2}.

  Quantum noise is one of the fundamental noise sources that limit the sensitivity of laser interferometric gravitational wave detectors \cite{Braginsky1996}. 
In the ground-based detectors, the quantum noise can be suppressed by using squeezed state of light \cite{GEO300, LIGOsqe, Virgosqe} and employing heavy mirrors. However, in the case of DECIGO, this strategy is not applicable. Using squeezed light in 1000-km-long arms results in too large a diffraction loss, and the mirror mass is limited by the satellite facility. Thus, we considered the quantum locking scheme\cite{qlock1, qlock2} to reduce quantum noise in DECIGO.

  In our earlier work on the quantum locking scheme \cite{qlock3}, we implemented, in simulation, the two short sub-cavities which share one mirror of the main cavity (Fig.\ref{cavity}). We found that the quantum noise can be optimized by taking an appropriate combination of output signals from the main cavity and the two sub-cavities. We also found that 
if we utilize the ponderomotive squeezing in the sub-cavities by sensing their length signals at an appropriate homodyne angle, we can reduce the quantum noise and even beat the standard quantum limit around 0.1 Hz. This frequency band is promising for detecting primordial gravitational waves.%

  Although the quantum locking scheme was found to be effective in reducing the quantum noise, it was found that we can reduce the quantum noise only in a relatively narrow frequency band. If we can reduce the quantum noise in a broader frequency band, the sensitivity of DECIGO to the primordial gravitational waves can be improved.

To reduce the quantum noise in a broader band, we consider detuning the sub-cavities from resonance to employ the optical spring effect\cite{opticalspring}. We expect that the larger optomechanical coupling and an additional adjustable parameter (detuning angle) provided by the optical spring could improve the quantum noise and could even broaden the frequency bandwidth of the quantum noise. To specify the optical spring effect in the quantum locking, it is important to numerically simulate the quantum noise in the quantum locking.

In this paper, first, we explain, in detail, a new method for reducing the quantum noise by using the quantum locking scheme with an added optical spring. Then we show, through simulation, how the quantum noise is reduced, and how the signal-to-noise ratio of the primordial gravitational wave for the quantum noise is improved.

\section{Theory}
\label{theo}
\begin{figure}[H]
\includegraphics[width=120mm,bb= 0 0 926 775]{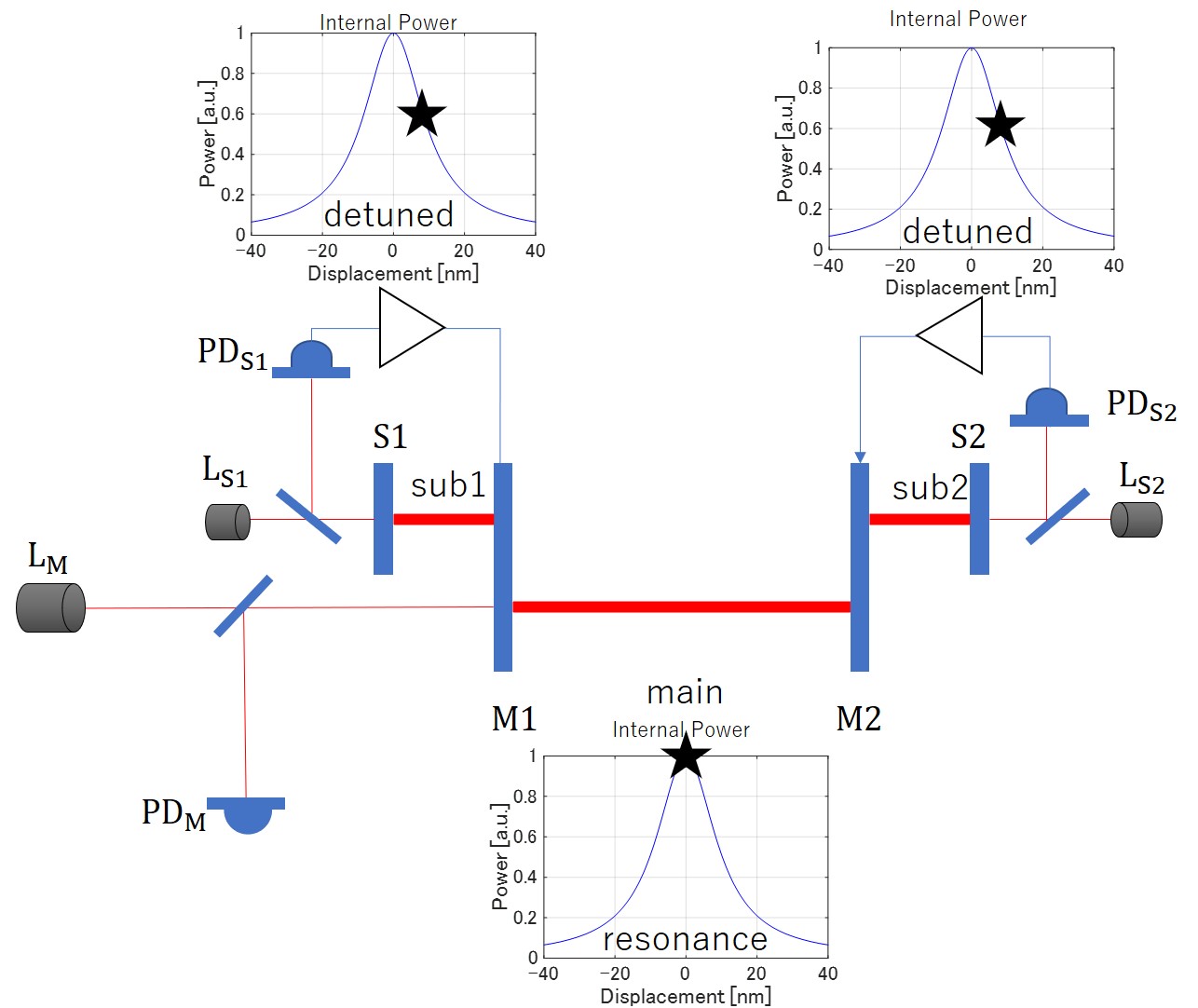}
\caption{Configuration of the quantum locking. M1 and M2 mirrors constitute the main cavity. Laser ($\rm L_M$) emits the light into the main cavity, and the reflected light is detected by a photodetector ($\rm PD_M$). The main cavity is controlled on resonance (marked by a "star" in the figure). Two sub-cavities consist of shared mirrors ($\rm M_1, M_2$) and additional mirrors ($\rm S_1, S_2$). They have their own lasers ($\rm L_{S1}, L_{S2}$) and photodetectors ($\rm PD_{S1}, PD_{S2}$). The sub-cavities are detuned from resonance (marked by ``star"s in the figure). }
\label{cavity}
\end{figure}

As shown in Fig.\ref{cavity}, in the quantum locking, we use sub-cavities which share mirrors with the main cavity. Let us name these sub-cavity sub-cavity1 and sub-cavity2.

In the quantum locking scheme, we obtain three output signals from the main cavity and sub-cavities. $V_0$ is the output signal from the main cavity, $V_1$ is that from sub-cavity1, and $V_2$ is that from sub-cavity2. Using these three output signals, we estimate the optimized output of the quantum locking scheme: $V$. If sub-cavity1 and sub-cavity2 have the same configuration as each other, the appropriate combination of these three output signals can be obtained by

\begin{equation}
V = V_0 + \chi \left( V_{1} + V_{2} \right),
\label{V}
\end{equation}

\noindent where $\chi$ is tunable function. Note that we have considered the above expression in the Laplace domain.  We can arbitrarily set $\chi$ to optimize the quantum noise.\cite{qlock3}.

We can beat the standard quantum limit if we use ponderomotive squeezing and homodyne detection in the sub-cavities.

In this paper, we use the quadrature-phase amplitude to describe quantum fluctuation \cite{two-phot}. 
We consider annihilation and creation operators of each cavity mode, $a_i$ and $a_i^\dagger$, which satisfy $[a_i, a_i^\dagger]=1$. We define $q_i=\frac{a_i + a_i^\dagger}{2}$ and $p_i=\frac{ a_i - a_i^\dagger}{2i}$
. $q_i$ is the amplitude quadrature and $p_i$ is the phase quadrature.  Here, $a_0$ is for the main cavity, $a_1$ is for the sub-cavity1 and $a_2$ is for the sub-cavity2.
$q_{i, \rm in}$ and $p_{i, \rm in}$ are amplitude and phase quadratures, respectively, of the incoming vacuum field of each cavity. $q_{i, \rm out}$ and $p_{i, \rm out}$ are those of the outgoing field.

\begin{figure}
\includegraphics[width=120mm,bb=  0 0 414 453]{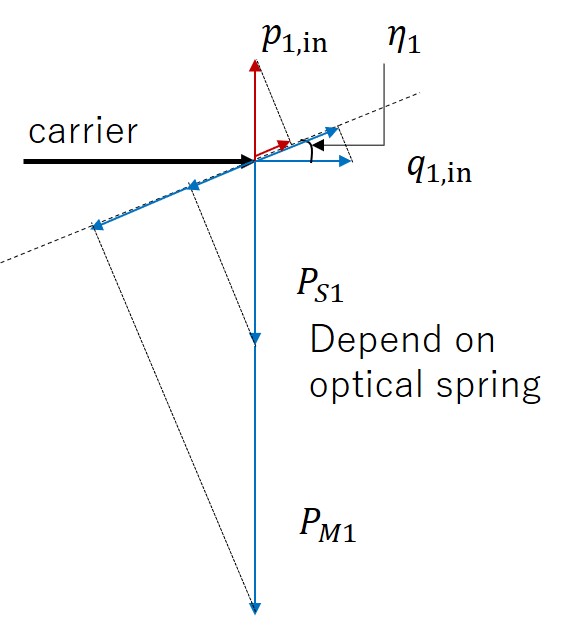}
\caption{Phasor diagram at the detection port of the sub-cavity. “carrier” is the laser light of sub-cavity 1. “$q_{1, \rm in}$” and “$p_{1,\rm in}$” are the amplitude and phase quadratures of the quantum fluctuations respectively. The amplitude quadrature combined with carrier shakes the mirror. “$P_{\rm S1}$” and “$P_{\rm M1}$” are the phase fluctuation due to the mirror displacement fluctuations, and their amplitudes depend on the optical spring effect. “$\eta_1$” is the homodyne angle. In homodyne detection, we detect the signals that are projected on the dotted line. }
\label{phase}
\end{figure}

 Figure \ref{phase} shows the phasor diagram at the detection port of the sub-cavity1. When the laser light enters the sub-cavity, the quantum fluctuations of the amplitude quadrature ($q_{1, \rm in}$) and the phase quadrature ($p_{1, \rm in}$) also enter the sub-cavity1. The amplitude quantum fluctuation couples with the carrier light and shakes the cavity mirrors. Then, the mirror displacement fluctuation causes phase fluctuations in the reflected light ($P_{\rm S1}, P_{\rm M1}$). If we detect the light along an appropriate axis (dotted line shown in Fig.\ref{phase}) by homodyne detection, we can cancel the phase fluctuation caused by the S1 mirror displacement fluctuation ($P_{\rm S1}$) and the amplitude quantum fluctuation ($q_{1, \rm in}$) at a certain frequency. It means that only the phase fluctuation caused by the M1 mirror displacement fluctuation ($P_{\rm M1}$) is detected at the photodetector. Thus, if we feed the signals back to the M1 mirror, we can eliminate the radiation pressure noise of the M1 mirror at a certain frequency.

\begin{figure}[H]
\includegraphics[width=120mm,bb=   0 0 533 690]{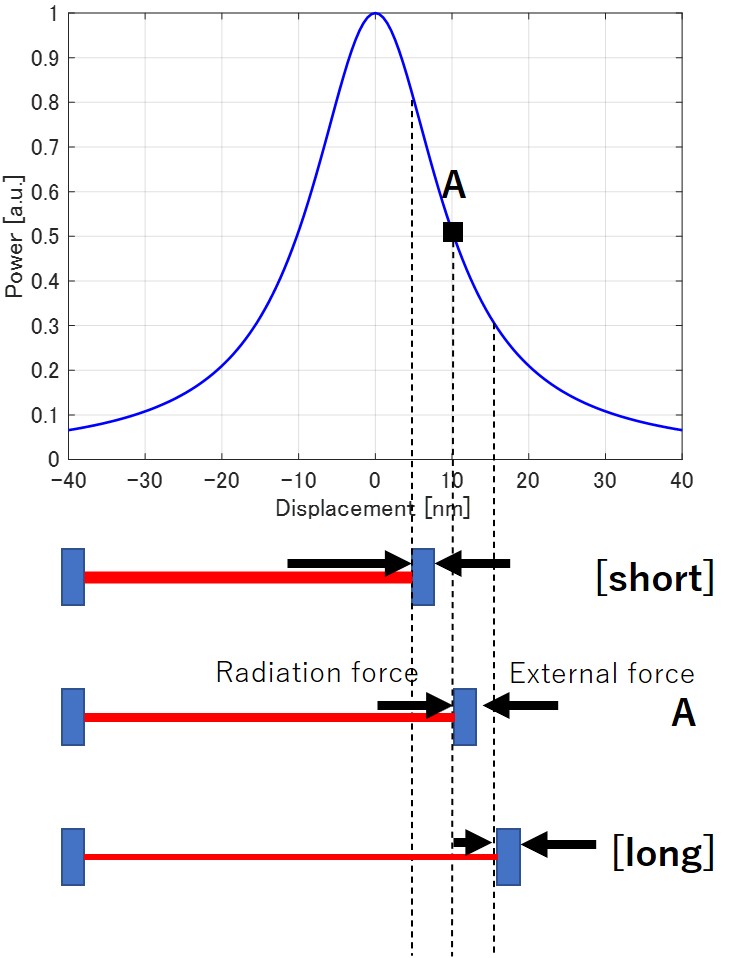}
\caption{Mechanism of the optical spring. The upper graph shows the internal power of the cavity vs. mirror displacement. When displacement is 0, the cavity is tuned exactly to resonance. When we detune the cavity from resonance (``A'' in the figure), the radiation force and the external force should be balanced. As a result, if the mirror moves and the cavity length decreases ([short] in the figure) or increases ([long] in the figure), the radiation force increases or decreases, respectively, and the mirror is pushed or pulled back.}
\label{opticalspring}
\end{figure}

 In this paper, additionally, we detune the sub-cavity from resonance and introduce an optical spring. The optical spring effect is caused in the detuned cavity. Generally, in a cavity, the radiation force acts on the cavity mirrors from the inside. To make the cavity stable, a constant external force that balances the radiation force is applied from the outside by the control system. In a detuned cavity, the radiation force depends on the length of the cavity. For example, in the cavity with a mirror placed initially on the declining slope (``A'' in Fig.\ref{opticalspring}), if the length of the cavity decreases ([short] in Fig.\ref{opticalspring}), the mirror is pushed back to the initial position by the increased radiation force, while if the length increases ([long] in Fig.\ref{opticalspring}), it is pulled back by the decreased radiation force. This is the optical spring.

\section{Simulation}
\subsection{Simulation model}
\label{Sim1}

\begin{figure}
\includegraphics[width=110mm,bb= 0 0 679 867]{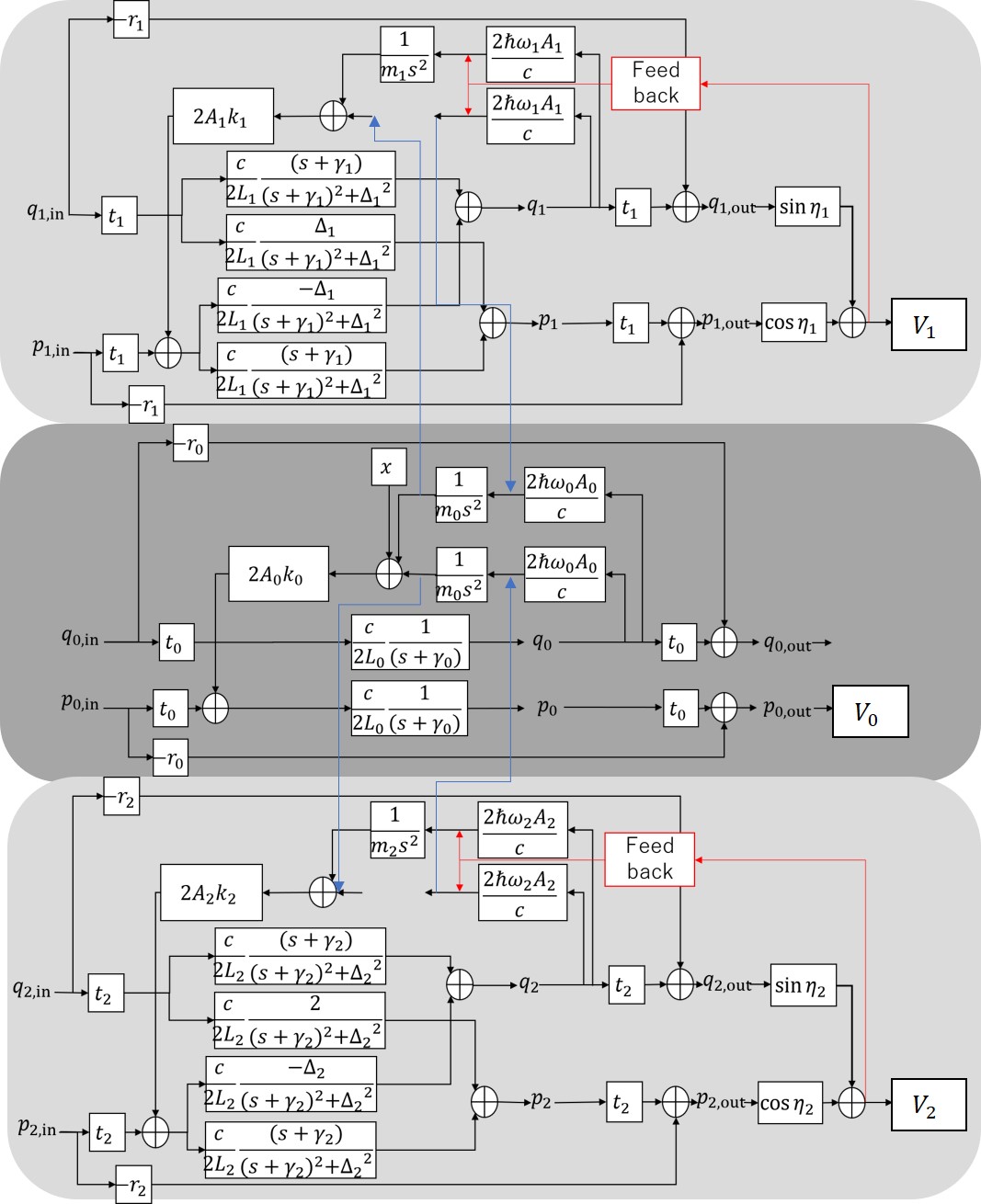}
\caption{Block diagram for calculating the quantum noise of the quantum locking with the optical spring. The central part is the main cavity, and the upper and lower parts are the sub-cavities. $q_{i, \rm in}$ and $p_{i, \rm in}$ are divided into reflection and transmission by $t_i$ and $r_i$. In the main cavity the transmission depends on the cavity pole: $\frac{c}{2L_0(s+\gamma_0)}$. $q_i$ and $p_i$ are amplitude and phase quadratures of the inter-cavity field. In the sub-cavities, the transmission is depends on the cavity pole and optical spring: $\frac{c(s+\gamma_1)}{2L_1\{(s+\gamma_1)^2+\Delta_1^2\}}$ or $\frac{\pm c \Delta_1}{2L_1\{(s+\gamma_1)^2+\Delta_1^2\}}$. Multiplied by $\frac{2 \hbar \omega_0 A_M}{c}$, the amplitude quadrature combines with the carrier and becomes the force pushing the mirror. Multiplied by $\frac{1}{ms^2}$, the force causes the displacement of the mirror. Further multiplied $2A_i k_i$, the mirror displacement is applied to the phase fluctuation. $q_{i, \rm out}$ and $p_{i, \rm out}$ are amplitude and phase quadratures of the outgoing field. The homodyne detector projects the signal into $\sin \eta_i$ and $\cos \eta_i$. Finally, we detect $V_i$. $x$ is the input port of gravitational wave signals indicated as the mirror displacement. }
\label{block}
\end{figure}

In order to calculate the quantum noise in the quantum locking scheme with the optical spring, we use a block diagram, shown in Fig.\ref{block} (see also \cite{EnomotoMron}). This block diagram is composed of three areas (gray areas in Fig.\ref{block}) representing the main cavity and the sub-cavities. Each cavity has two input ports, the amplitude quadrature, $q_{i, \rm in}$ ($i=0,1,2$; 0 is for the main cavity, 1 and 2 are for the sub-cavities), and the phase quadrature, $p_{i, \rm in}$, and one output port ($V_i$).  
Note that, in this block diagram, we assume that each cavity is over-coupled. 

In the main cavity, the amplitude quadrature ($q_{0,\rm in}$) and the phase quadrature ($p_{0, \rm in}$) are divided into transmission and reflection according to the input mirror's amplitude transmissivity, $t_0$, and its amplitude reflectivity, $r_0$, respectively. Here, we assume that the mirrors have no loss: $t_i^2+r_i^2=1$. After that, the amount of transmitted light depends on the cavity pole in the cavity: $\frac{c}{2L_0(s+\gamma_0)}$, where $s$ is the Laplace complex variable and $\gamma_i$ is the cavity pole. 

\begin{eqnarray}
\gamma_i &=& \frac{\pi c}{2 L_i {\mathcal F}_i} \\
{\mathcal F}_i &=& \frac{\pi \sqrt{r_i}}{1-r_i}.
\end{eqnarray}

\noindent Here, $c$ is the speed of light and $L_i$ is the cavity length. Within the cavity, the amplitude quadrature and the phase quadrature are represented by $q_0$ and $p_0$. The amplitude quadrature couples with carrier light and becomes a force that pushes the mirror by $\frac{2 \hbar \omega_0 A_0}{c}$, where $\hbar$ is the reduced Planck constant, $\omega_i$ is the angular frequency of the light: $\omega_i = \frac{2 \pi c}{\lambda_i}$, and $A_i$ is the amplitude of the light: $A_i=\frac{2 I_i}{\omega_i \hbar}$, where $I_i$ is the intensity of the light. The force causes the mirror to displace by $\frac{1}{m_0s^2}$, where $m_i$ is the mirror mass. The mirror displacement is multiplied by $2A_0 k_0$and added the phase fluctuation, where $k_i$ is the wavenumber of laser light: $k_i=\frac{\omega_i}{c}$. The quantum fluctuations go out the main cavity; $t_0$, and they are represented by $q_{0, \rm out}$ and $p_{0, \rm out}$. Finally, we detect the signal as $V_0$. $x$ is the input port of gravitational wave signals as mirror displacement.

In the sub-cavity1, the transmitted fluctuations are redistributed into the amplitude quadrature and the phase quadratures: $\frac{c(s+\gamma_1)}{2L_1\{(s+\gamma_1)^2+\Delta_1^2\}}$ or $\frac{\pm c \Delta_1}{2L_1\{(s+\gamma_1)^2+\Delta_1^2\}}$.

\begin{equation}
\Delta_i = \frac{\delta \phi_i c}{2L_i}.
\end{equation}

\noindent Here, $\delta \phi_i$ is the detuning angle. After that, the new amplitude quadrature shakes the mirrors and is added to the phase quadrature in the same manner as the main cavity. Note that since the main cavity and the sub-cavity share their mirrors, they also share block: $\frac{1}{m_0s^2}$. $V_1$ is obtained through homodyne detection, which is represented by $\sin \eta_1$ and $\cos \eta_1$. The block diagram of the sub-cavity2 is the same as that of the sub-cavity1.

Using this block diagram, we calculate the optimized quantum noise. First, we obtain each photodetector's signals as follows:

\begin{eqnarray}
V_0 &=& s + A q_{0, \rm in} + i B p_{0, \rm in} + C q_{1, \rm in} + i D p_{1, \rm in} + E q_{2, \rm in} + i F p_{2, \rm in}  \\
V_1 &=& G q_{0, \rm in} + i H p_{0, \rm in} + I q_{1, \rm in} + i J p_{1, \rm in}  \\
V_2 &=& K q_{0, \rm in} + i L p_{0, \rm in} + M q_{2, \rm in} + i N p_{2, \rm in}. 
\label{Vs2}
\end{eqnarray}

\noindent Here, A through N are the coefficients for each independent noise source.

Then we take the appropriate combination of $V_0$, $V_1$ and $V_2$ according to eq.\ref{V}. To estimate the power spectrum density, we take a quadrature sum of the contributions to $V$ from each independent noise source. The power spectral density of the total noise is minimized when 

\begin{equation}
\chi = - \frac{2(AG^* + BH^* + CI^* + DJ^*)}{ \left( |2G|^2 + |2H|^2 +2 |I|^2 + 2|J|^2 \right)}.
\end {equation}

\noindent A detailed calculation can be found in \cite{qlock3}

\subsection{Simulation conditions}
\label{Sim2}

In this subsection, we state the parameters used to estimate the noise power spectral density and the signal-to-noise ratio (SNR) for the primordial gravitational waves in this paper. 

Table \ref{param1} shows the parameters for the block diagram used to estimate the noise power spectral density.  In our simulation, we consider the case where sub-cavity2 has the same configuration as sub-cavity1.

\begin{table}[htb]
\caption{Parameters for the block diagram. }
  \begin{tabular}{|l|c|c|r|} \hline
    Main cavity & Laser power & $I_0$ & 100 W  \\ \cline{2-4}
      & Finesse &${\mathcal F}_0$ & 10  \\ \cline{2-4}
      & Cavity length & $L_0$ &1000 km  \\ \cline{2-4}
      & Wavelength & $\lambda_0$ &515 nm  \\ \cline{2-4}
      & Mirror mass & $m_0$ & 100 kg  \\ \hline
    Sub cavity  & Laser power &$I_1, I_2$& *  \\ \cline{2-4}
      & Finesse &${\mathcal F}_1, {\mathcal F}_2$ & *  \\ \cline{2-4}
      & Cavity length & $L_1, L_2$ & 1 m  \\ \cline{2-4}
      & Wavelength & $\lambda_1, \lambda_2$ & 515 nm  \\ \cline{2-4}
      & Mirror mass & $m_1, m_2$ &100 kg  \\ \cline{2-4}
      & Homodyne angle& $\eta_1, \eta_2$ & Free \\ \cline{2-4}
      & Detuning angle & $\delta\phi_1, \delta\phi_2$& Free \\ \hline
  \end{tabular}
\label{param1}
\end{table}

\noindent the homodyne and detuning angles are free parameters. In Sec.\ref{Simr1}, we fix the finesse at 10 and the laser power at 100 W in the sub-cavities. After that, in Sec.\ref{Sim3},  we regard the finesse and laser power of the sub-cavities are free parameters. Note that we limit the laser power to 100 W keeping practical constrain in mind.

We calculate the signal-to-noise ratio (SNR) for the primordial gravitational waves and optimize the homodyne and detuning angles. To calculate the SNR, we use the following equation \cite{SNRequation};

\begin{equation}
SNR = \frac{3 H_0^2}{10 \pi^2} \sqrt{T} \left[\int_{0.1}^{1} df  \frac{2 \Gamma(f)^2 \Omega_{GW}^2(f)}{f^6 P_1(f) P_2(f)} \right]^{1/2}.
\label{snr_eq}
\end{equation}

\noindent Here, $P_1$ and $P_2$ are the noise power spectrum densities calculated in Sec.\ref{Sim1} with Table.\ref{param1}. 
$H_0$ is the Hubble parameter, $T$ is the correlation time and $\Omega_{\rm GW}$ is the energy density ratio of the primordial gravitational wave to the closure density \cite{Planck2018, Kuroyanagi2014}.
We integrate the quantity in the frequency space from 0.1 Hz to 1 Hz, which is the target frequency band of DECIGO. 
$\Gamma$ is the correlation function, in the case of DECIGO, $\Gamma=1$.
Table \ref{parasnr} shows the actual numbers used in the calculation.

\begin{table}[htb]
\begin{center}
\caption{Parameters  used to estimate the SNR.}
\begin{tabular}{|c|c|c|}\hline
Noise  power spectral densities &$P_1, P_2$ & calculeted in sec.\ref{Sim1} \\ \hline
 Hubble parameter & $H_0$ & 70 ${\rm km \cdot sec^{-1} \cdot Mpc^{-1}} $  \\ \hline
Time for correlation &$T$ & 3 years \\ \hline
Energy density & $\Omega_{\rm GW}$  &$10^{-16}$  \\ \hline
Frequency &$f$ & 0.1 to 1 Hz \\ \hline
correlation function &$\Gamma$  &1  \\ \hline
\end{tabular}
\label{parasnr}
\end{center}
\end{table}

\subsection{Dependence of signal-to-noise ratio on homodyne angle and detuning angle}
\label{Simr1}

\begin{figure}[H]
\includegraphics[width=100mm,bb=  0 0 420 315]{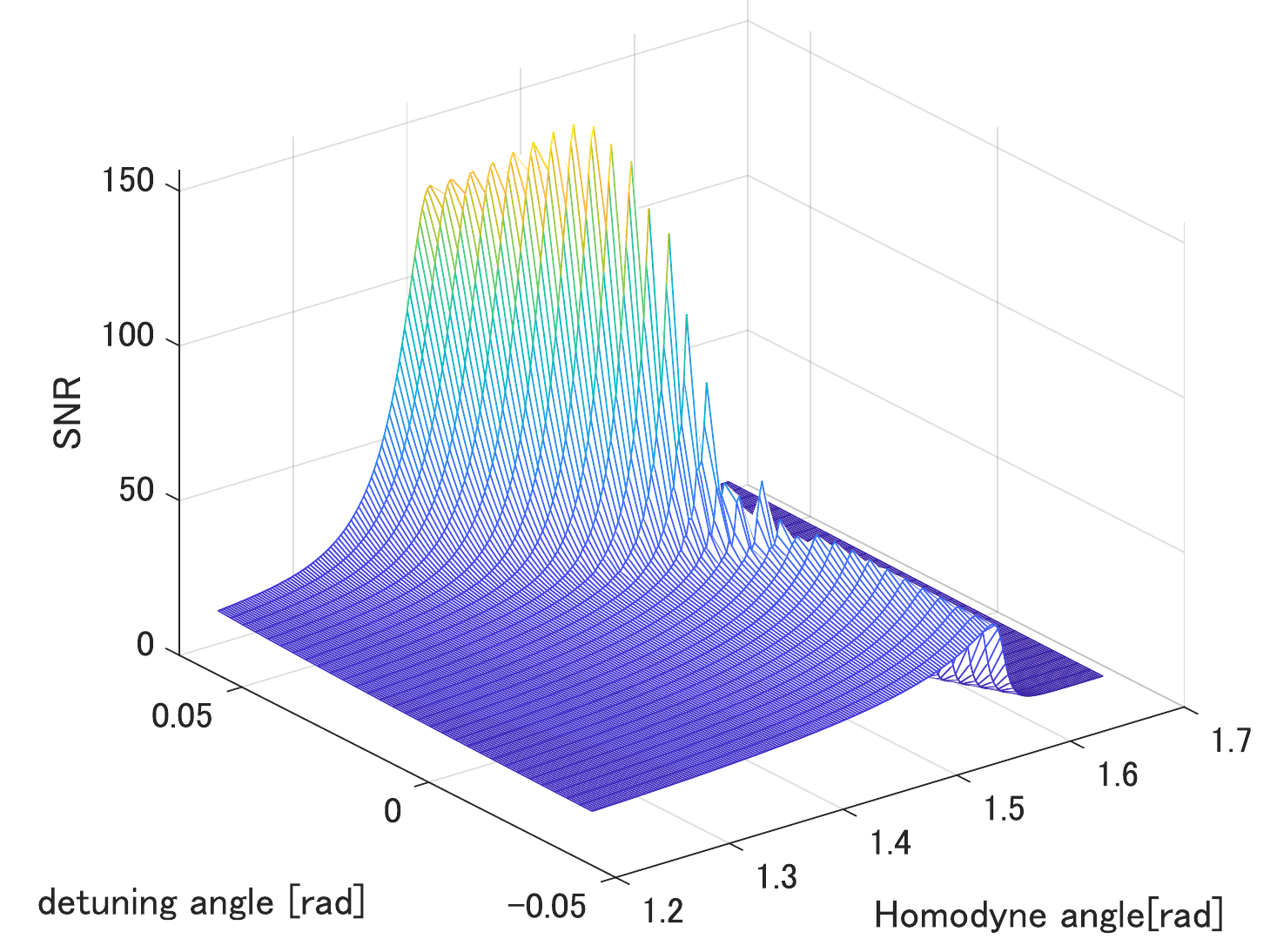}
\caption{Dependence of SNR on the homodyne and detuning angles. The best SNR is 156.7 when the detuning angle is 0.04 rad and the homodyne angle is 1.477 rad. The rugged features on the ridge in the curved surface are caused by the imperfect resolution of the detuning angle and the homodyne angle in the calculation.
}
\label{threeD}
\end{figure}

Figure \ref{threeD} shows the simulation result of the dependence of SNR on the homodyne and detuning angles when we fix the finesse to 10 and the laser power to 100 W in the sub-cavities. 

We show the optimal homodyne angle and the detuning angle in Fig.\ref{threeD}.
When the sub-cavities are off resonance, the best SNR is 156.7. On the other hand, when the sub-cavities are on resonance (which means $\delta \phi_i$ is 0), the best SNR is 42.2.
The off-resonant sub-cavities with an optical spring provide an improvement which is factor of 3.7 better than the resonant sub-cavities in SNR.

\begin{figure}[H]
\includegraphics[width=100mm,bb=  0 0 420 315]{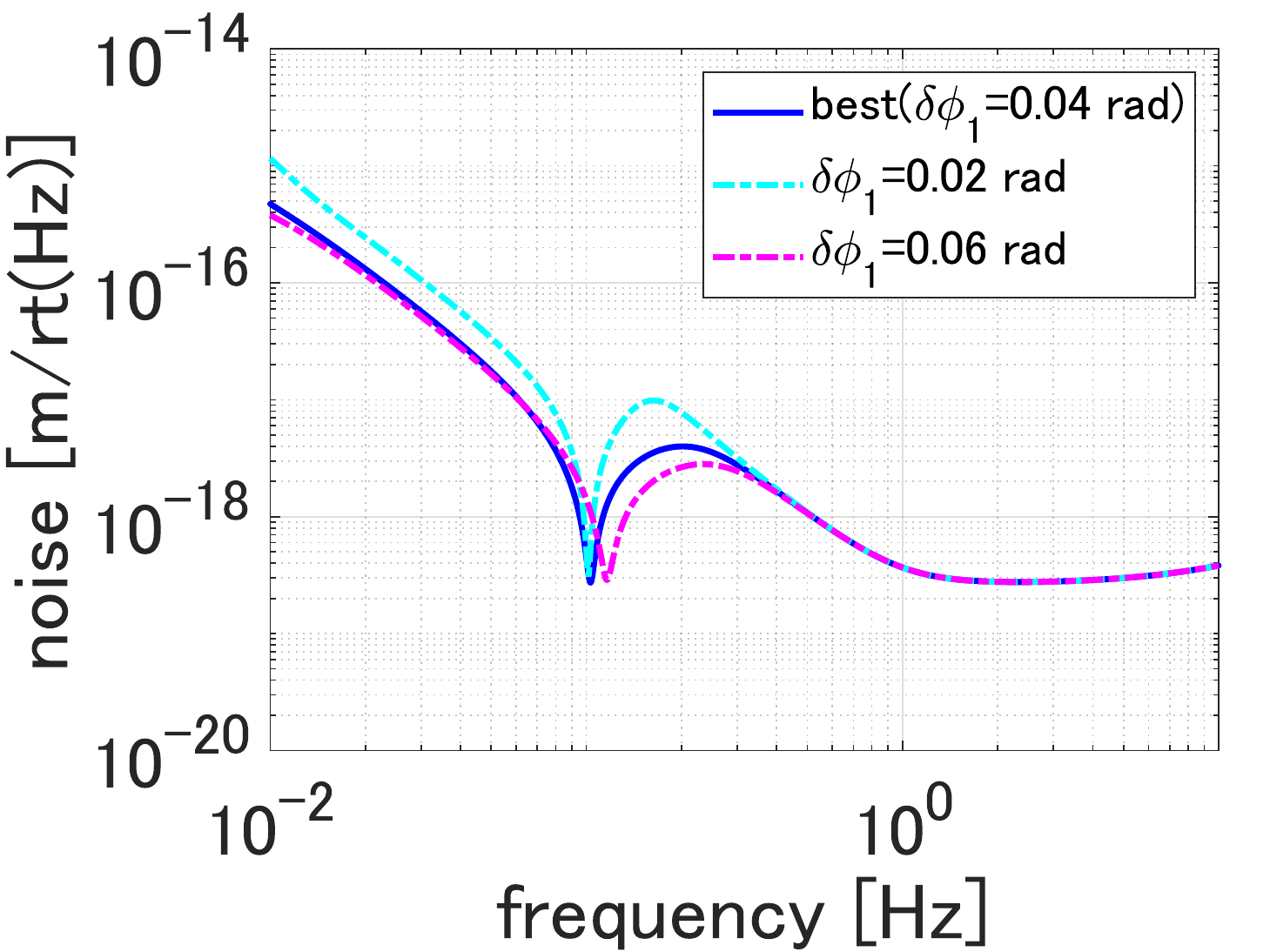}
\caption{Figure: Sensitivity curves at three points on the ridge in the curved surface in Fig.\ref{threeD}. The blue curve provides the best SNR with an optimized detuning angle ($\delta \phi_1=0.04$ rad). The cyan curve is for a smaller detuning angle  ($\delta \phi_1=0.02$ rad), and the magenta curve is for a larger detuning angle ($\delta \phi_1=0.06$ rad).
}
\label{total10100}
\end{figure}

Figure.\ref{total10100} shows the sensitivity curves at three points on the ridge in the curved surface in Fig.\ref{threeD}. For the detuning angle smaller than the best-SNR detuning angle, the quantum noise is reduced in a narrower frequency band. On the other hand, for the larger detuning angle the dip frequency is moved to a higher frequency.


\subsection{Dependence of signal-to-noise ratio on finesse and laser power}
\label{Sim3}

\begin{figure}[H]
\includegraphics[width=100mm, bb=0 0 762 466]{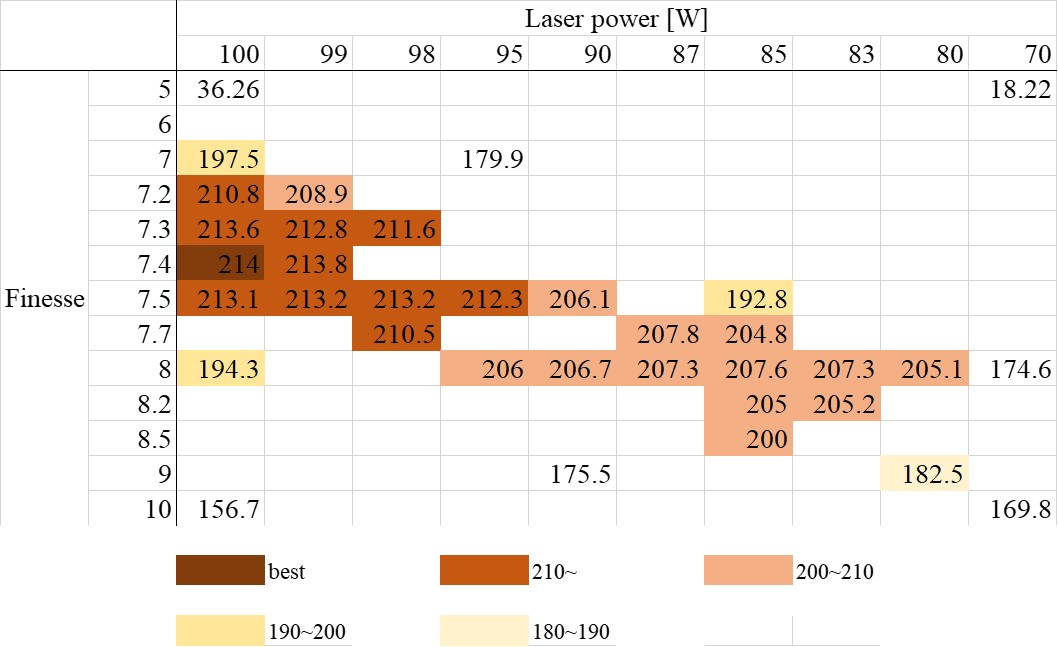}
\caption{Dependence of SNR(${\mathcal F}, I$) on the finesse and laser power of sub-cavities. The best SNR is 214, when finesse is 7.4 and laser power is 100 W. The blank indicates that the SNR is not calculated as it is not expected to be high.}
\label{SNR_FI}
\end{figure}

In this subsection, we show the simulation results under the condition that the finesse and the laser power of the sub-cavities are also regarded as free parameters. For each pair of finesse and laser power, ${\mathcal F}_1$  and $I_1$, we optimize the homodyne angle ($\eta_1$) and the detuning angle ($\delta\phi_1$) to make SNR the highest. Here, we define this highest SNR as SNR(${\mathcal F}, I$).
Figure \ref{SNR_FI} shows the dependence of SNR(${\mathcal F}, I$) on ${\mathcal F}_1$ and $I_1$. And the best SNR(${\mathcal F}, I$) is 214, when finesse is 7.4 and laser power is 100 W.

In order to compare this result with the resonant sub-cavities case, we performed the same calculation for the resonant case. It was found that the best SNR(${\mathcal F}, I$) is 84.8,  when finesse is 171.3 and laser power is 100 W. Note that we put “for example” because in the resonant case, the SNR is the same if the product of finesse and laser power is the same. 

\begin{figure}[H]
\includegraphics[width=90mm,bb=  0 0 420 315]{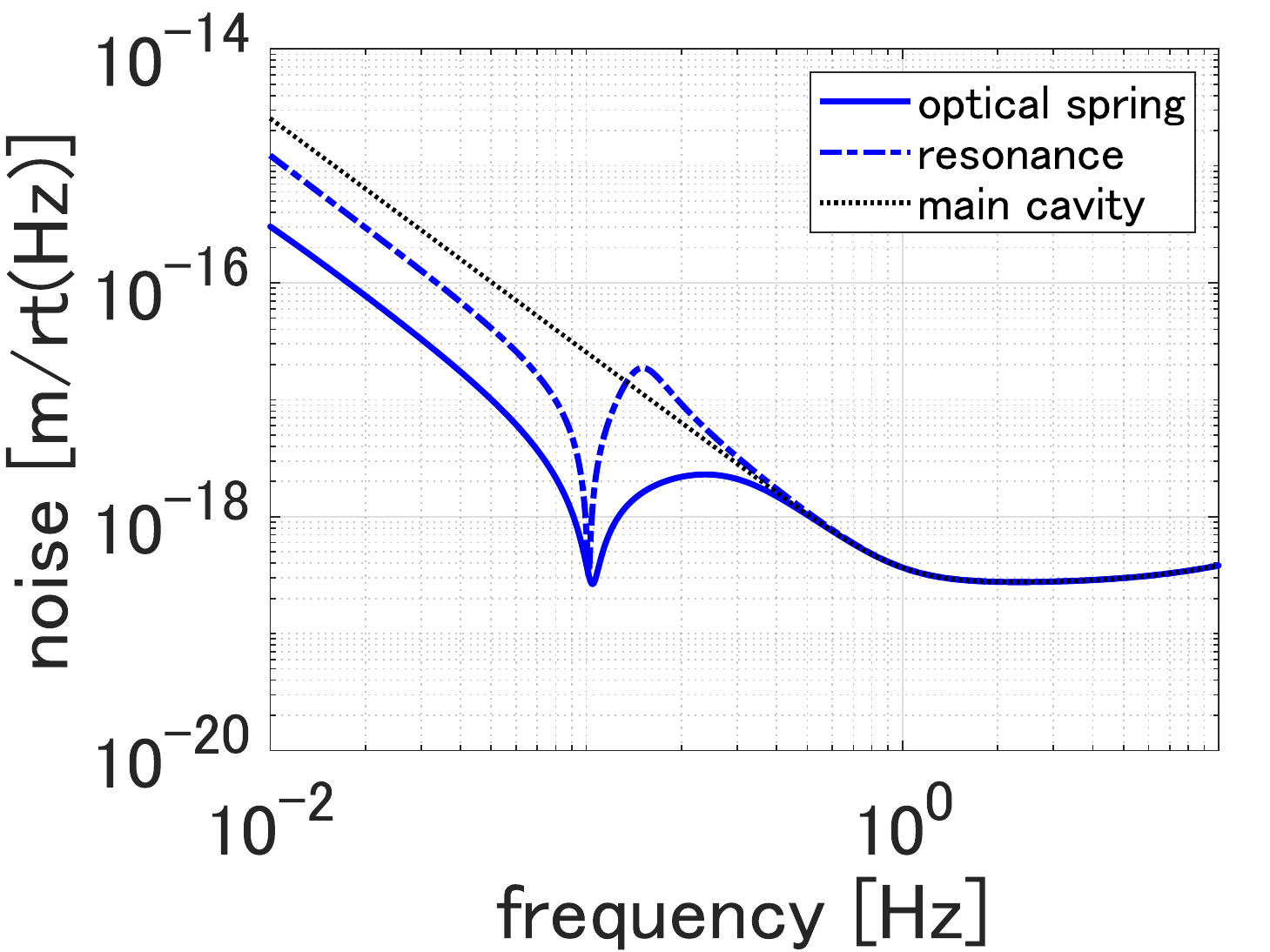}
\caption{Comparison of total noise curve with off-resonant sub-cavities and resonant sub-cavities. The black dotted line shows the total noise curve without sub-cavities as reference.
}
\label{total}
\end{figure}

Figure \ref{total} shows the total noise curves for the highest SNR(${\mathcal F}, I$) with the resonant sub-cavities and off-resonant sub-cavities. 

\noindent The total noise with the off-resonant sub-cavities (blue dot line in Fig.\ref{total}) is reduced in a broader frequency band than with the resonant sub-cavities (blue line in Fig.\ref{total}). The dip in the total noise curve with the off-resonant sub-cavities is deeper than that with the resonant sub-cavities. Because the primordial gravitational wave signal is larger at lower frequencies, the optimized total noise curve has a dip around 0.1 Hz, which is the lowest frequency of the integration frequency range, where the best SNR is obtained.

\section{Discussion}

In this section, we discuss the reason for the improvement in SNR. 

\begin{figure}[htb]
\begin{tabular}{cc}
\begin{minipage}{0.5\hsize}
 \includegraphics[width=45mm, bb=0 0 420 315]{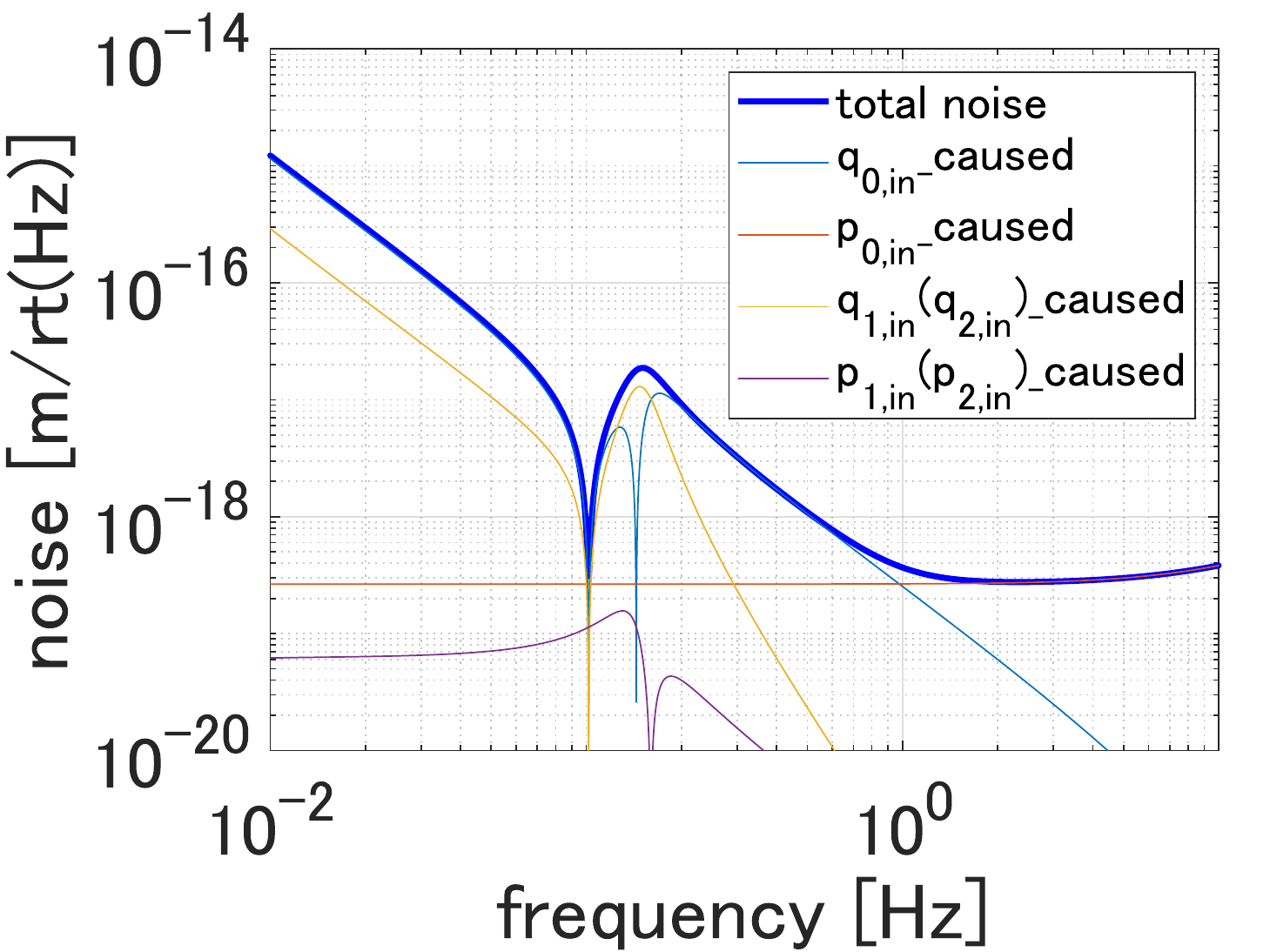}
 \subcaption{}
\label{SNRa}
\end{minipage}
\begin{minipage}{0.5\hsize}
 \centering
 \includegraphics[width=45mm, bb=0 0 420 315]{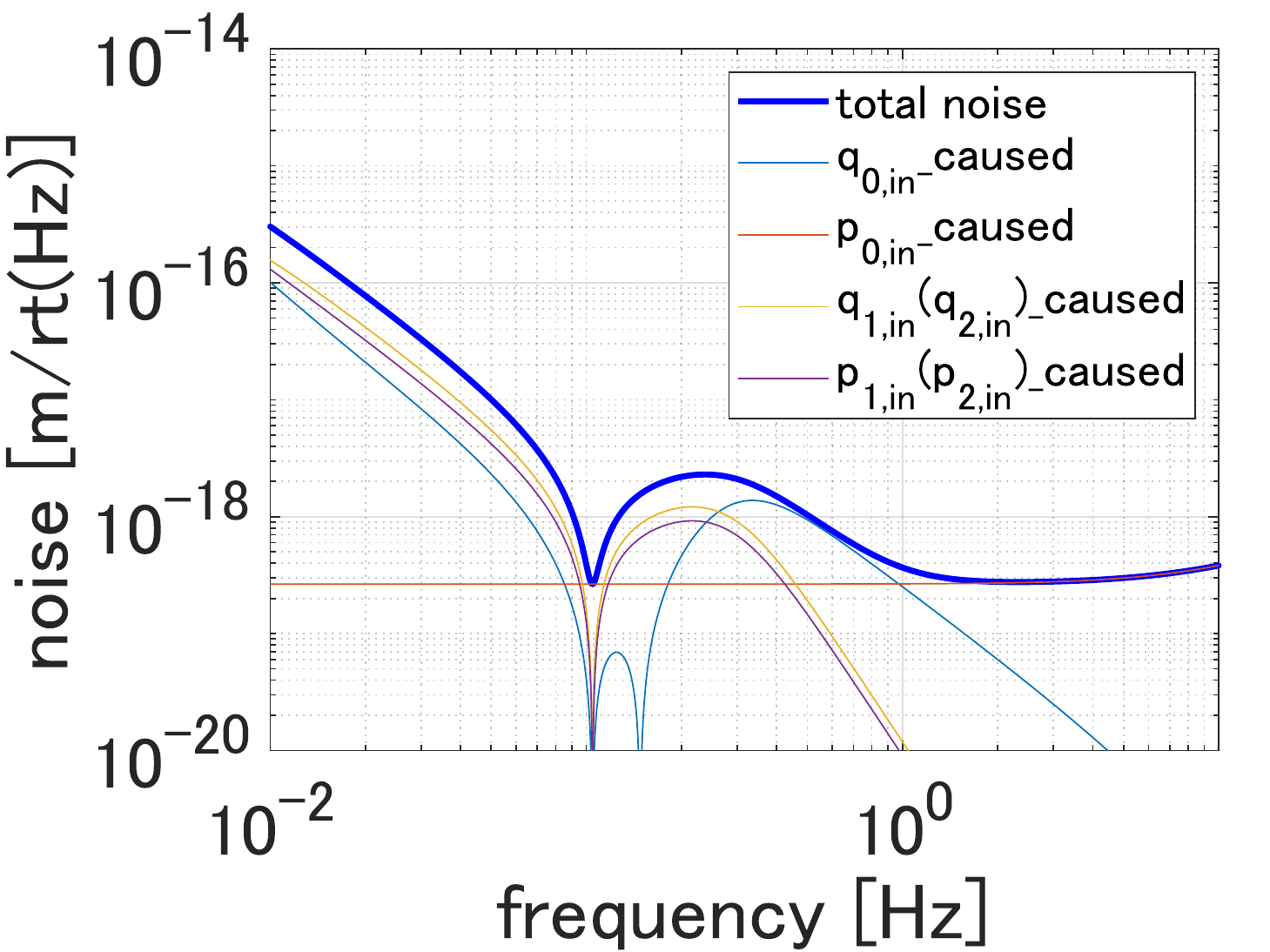}
 \subcaption{}
\label{SNRb}
\end{minipage}
\end{tabular}
\caption{Sensitivity curve for the best SNR(${\mathcal F}, I$) for the case with the resonant sub-cavities (\ref{SNRa}), and for the case with off-resonant sub-cavities (\ref{SNRb}). The noise budgets for each quantum noise are also plotted.}
\label{SNR}
\end{figure}

Figure \ref{SNR} shows the noise budget with the resonant sub-cavities (\ref{SNRa}) and off-resonant sub-cavities (\ref{SNRb}). Around the dip frequency, in the resonant sub-cavities case, $q_{0,\rm in}$\_caused noise and $q_{1,\rm in}$ ($q_{2,\rm in}$)\_caused noise are close to limiting the total quantum-noise sensitivity, while $p_{1,\rm in}$ ($p_{2,\rm in}$)\_caused noise is negligible. When we detune the sub-cavities from resonance, $q_{0,\rm in}$\_caused noise and $q_{1,\rm in}$ ($q_{2,\rm in}$)\_caused noise decrease at the expense of an increase in $p_{1,\rm in}$ ($p_{2,\rm in}$)\_caused noise. As a result, the total quantum noise with the off-resonant sub-cavities is reduced around the dip frequency. This improvement can be regarded as an optimizing shuffle of several quantum noises thanks to the additional optomechanical free parameter (detuning angle).

Incidentally, $p_{0,\rm in}$\_caused noise is not affected by the optical spring because this noise corresponds to the shot noise of the main cavity. This noise limits the depth of the dip in both cases.

We can also notice that for the resonant case, the dip frequency of  $p_{1,\rm in}$ ($p_{2,\rm in}$)\_caused noise is different from that of $q_{0,\rm in}$\_caused noise and $q_{1,\rm in}$ ($q_{2,\rm in}$)\_caused noise. On the other hand, for the off-resonant case, the dip frequencies of  $p_{1,\rm in}$ ($p_{2,\rm in}$)\_caused noise, $q_{0,\rm in}$\_caused noise, and $q_{1,\rm in}$ ($q_{2,\rm in}$)\_caused noise are all the same.


\begin{figure}
\begin{tabular}{cc}
\begin{minipage}{0.5\hsize}
 \includegraphics[width=45mm, bb=0 0 420 315]{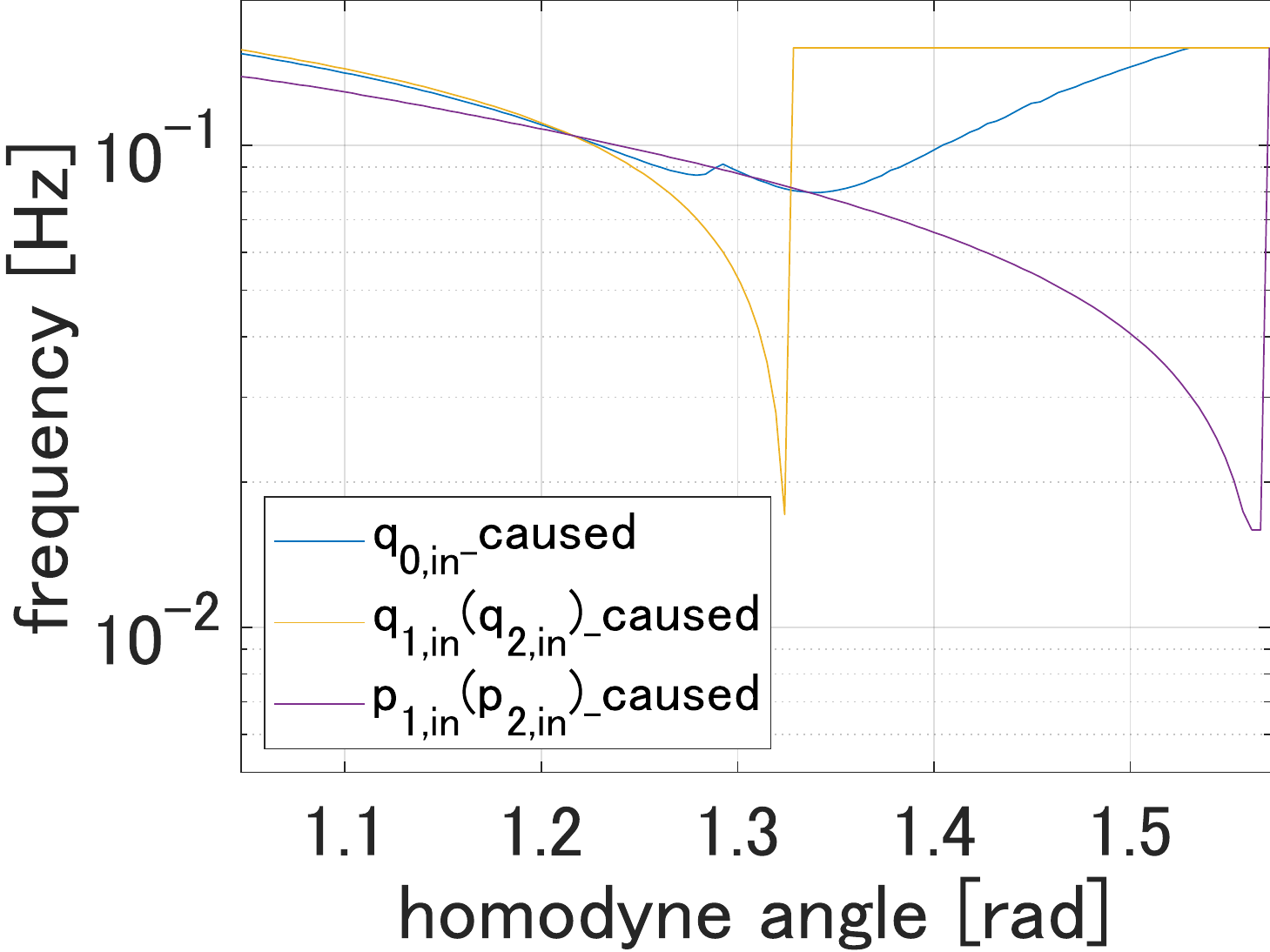}
 \subcaption{}
\label{freqe}
\end{minipage}
\begin{minipage}{0.5\hsize}
 \centering
 \includegraphics[width=45mm, bb=0 0 420 315]{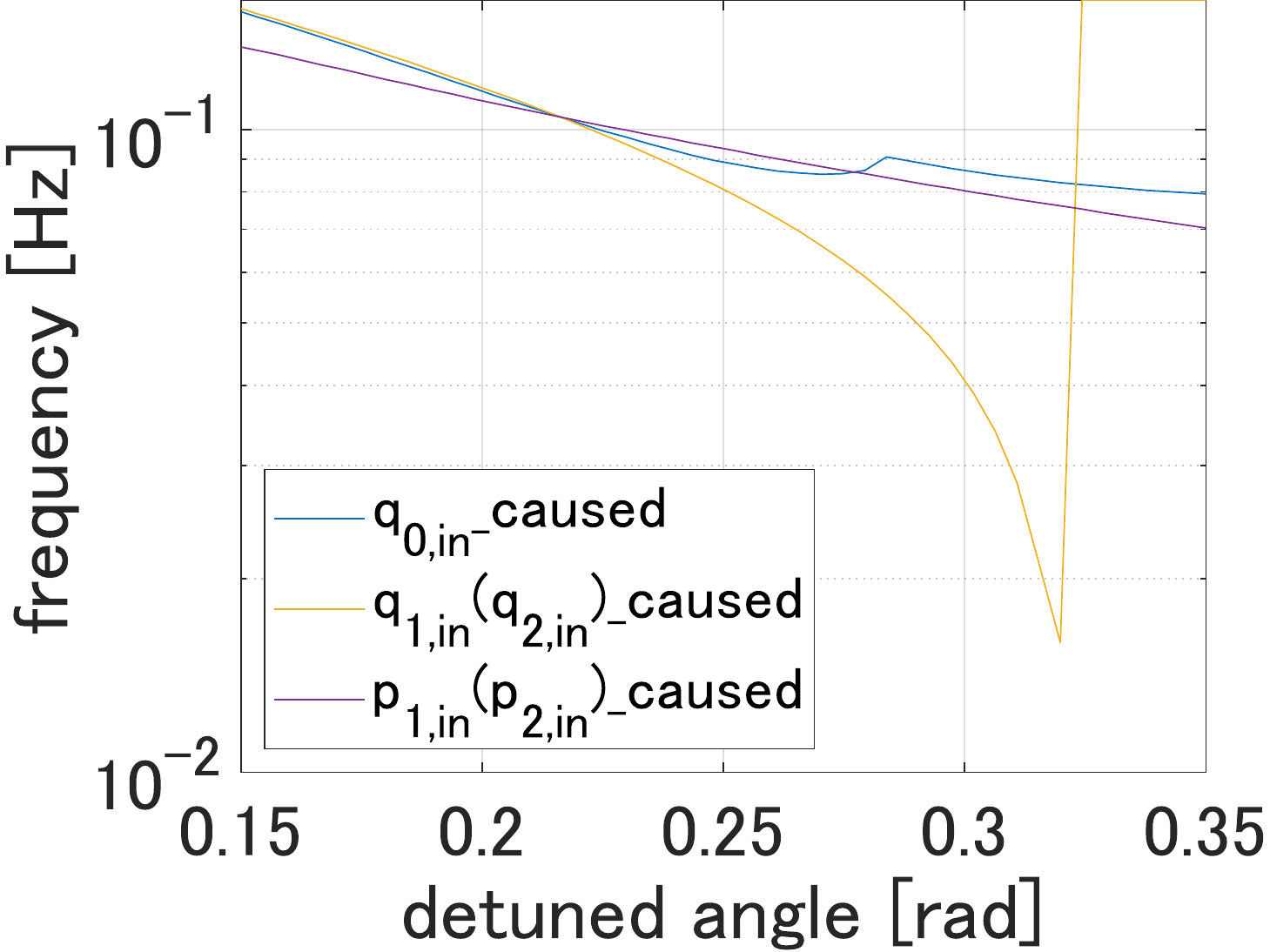}
 \subcaption{}
\label{freqp}
\end{minipage} \\
\begin{minipage}{0.5\hsize}
 \centering
 \includegraphics[width=45mm, bb=0 0 420 315]{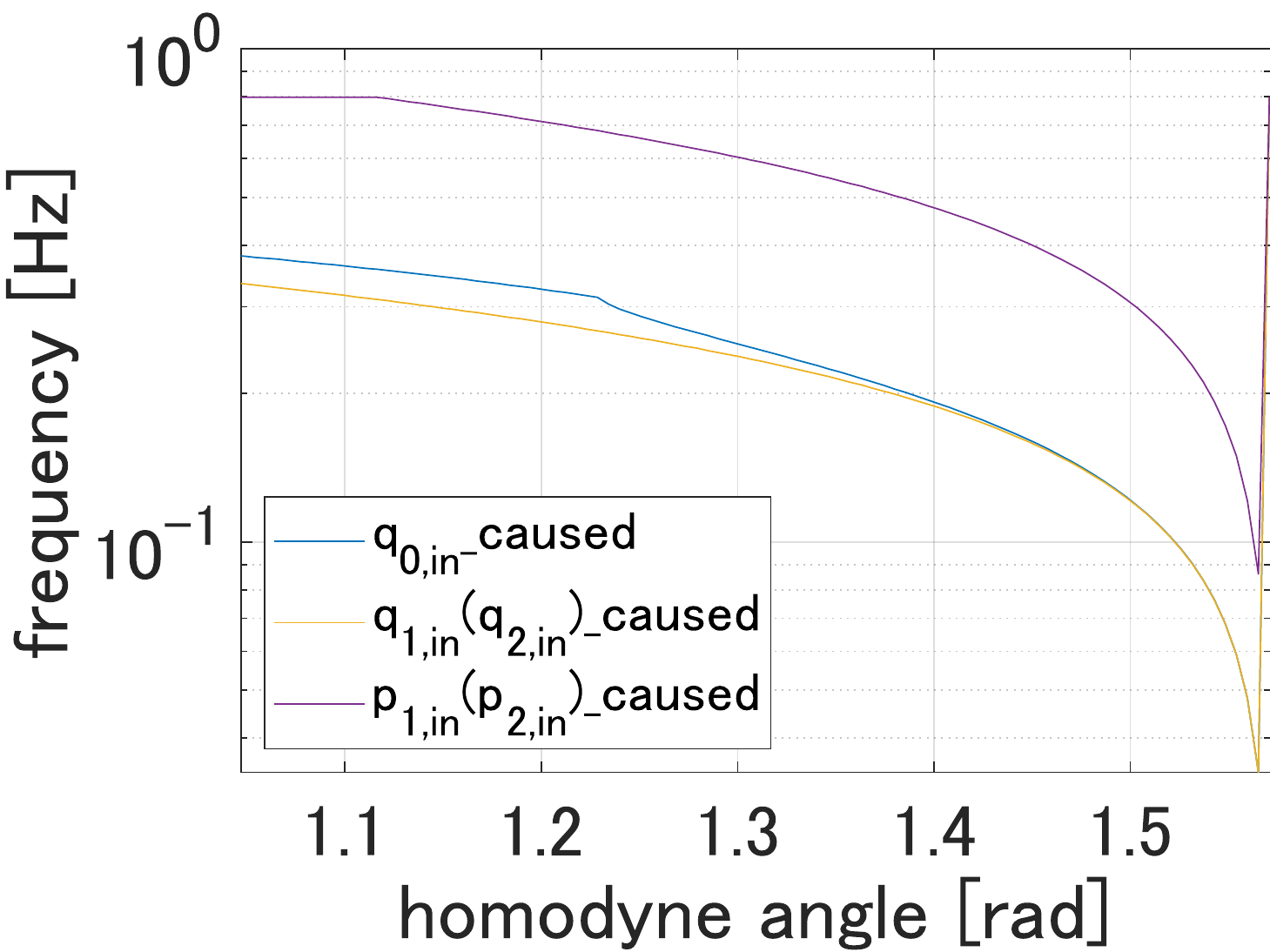}
 \subcaption{}
\label{freqe2}
\end{minipage}
\end{tabular}
\caption{Dependence of the dip frequencies of the three quantum noises on the homodyne angle and detuning angle with $({\mathcal F}_1, I_1)=(7.4, 100)$.  The off-resonant case is shown in (\ref{freqe}) and (\ref{freqp}). In (\ref{freqe}), the detuning angle is fixed at 1.216 rad, and in (\ref{freqp}), the homodyne angle is fixed at 0.216 rad. The resonant case is shown in (\ref{freqe2}), where the detuning angle is zero.}
\label{dipfreq}
\end{figure}

The dip frequencies of these three quantum noises are determined by the homodyne and detuning angles when we fix the finesse and the laser power of sub-cavities. Figure \ref{dipfreq} shows the dependence of the dip frequencies of the three quantum noises on the homodyne and detuning angles for the off-resonant case (\ref{freqe} and \ref{freqp}) and for the resonant case (\ref{freqe2}) with the parameters (${\mathcal F}_1, I_1$) = (7.4, 100), which provides the best SNR(${\mathcal F}, I$). In (\ref{freqe}) and (\ref{freqp}), these three dip frequencies cross at one frequency near 0.1 Hz for the particular pair of the homodyne angle and detuning angle. However, In (\ref{freqe2}), these three dip frequencies do not cross at one frequency. This difference can be attributed to the fact that the off-resonant case has the additional free parameter (detuning angle) to tune the dip frequencies of the three quantum noises.

\section{Conclusion}
Encouraged by the result of our previous work on a quantum locking scheme for DECIGO, in this paper, we explored the use of an optical spring in the sub-cavities of the quantum locking system, with expectation that enhanced optomechanical coupled would lead to improved sensitivity.
 We performed simulations with detuning included, and found that by optimizing detuning angle of sub-cavities, the total quantum noise is decreased in a broader frequency band compared with the resonant case. We also found that this improvement can be attributed to the shuffle of the three quantum noises as well as the adjustment of the dip frequencies of the three quantum noises thanks to the additional free parameter (detuning angle). We believe that this quantum locking scheme with an optical spring provides a promising technology that would enhance the reach of DECIGO.

\section*{Acknowledgments}
We would like to thank Keiko Kokeyama for helpful discussion.
We would like to thank Matthew Evans and Dhruva Ganapathy for English editing. 
This work was supported by JSPS KAKENHI Grant Number JP19H01924.





\end{document}